\def\apj{ApJ}
\def\mnras{MNRAS}

\def\km{{\rm\thinspace km}}

\def\Mpc{{\rm\thinspace Mpc}}

\def\s{{\rm\thinspace s}}

\def\kmps{\hbox{$\km\s^{-1}\,$}}

\def\kmpspMpc{\hbox{$\kmps\Mpc^{-1}$}}

\def\spose#1{\hbox to 0pt{#1\hss}}
\def\approxlt{{\mathrel{\spose{\lower 3pt\hbox{$\sim$}}
        \raise 2.0pt\hbox{$<$}}}}
\def\approxgt{\mathrel{\spose{\lower 3pt\hbox{$\sim$}}
        \raise 2.0pt\hbox{$>$}}}
\def\tento#1{\times 10^{#1}}
\newcommand\beq{\begin{equation}}
\newcommand\eeq{\end{equation}}
\makeatletter

\newenvironment{figurehere}
  {\def\@captype{figure}}
  {}
\makeatother

\documentstyle[emulateapj,psfig]{article}

\begin{document}

\medskip

\setcounter{footnote}{0}

\title{Radio Foregrounds for the 21cm Tomography of the Neutral
Intergalactic Medium at High Redshifts}

\author{Tiziana Di Matteo\altaffilmark{1,2}, Rosalba Perna\altaffilmark{1,3}, Tom 
Abel \altaffilmark{1,4}
and Martin J. Rees \altaffilmark{4}}

\altaffiltext{1}{Harvard Smithsonian Center for Astrophysics, 60 Garden Street, 
Cambridge, MA 02138}

\altaffiltext{2}{{\em Chandra\/} Fellow}

\altaffiltext{3}{Harvard Junior Fellow}

\altaffiltext{4}{Institute of Astronomy, Madingley Road, Cambridge, CB03 0HA, UK
}

\begin{abstract}

Absorption or emission against the cosmic microwave background
radiation (CMB) may be observed in the redshifted 21cm line if the
spin temperature of the neutral intergalactic medium prior to
reionization differs from the CMB temperature. This so-called 21cm
tomography should reveal important information on the physical state
of the intergalactic medium at high redshifts. The fluctuations in the
redshifted 21 cm, due to gas density inhomogeneities at early times,
should be observed at meter wavelengths by the next generation radio
telescopes such as the proposed {\it Square Kilometer Array (SKA)}.
Here we show that the extra-galactic radio sources provide a
serious contamination to the brightness temperature fluctuations
expected in the redshifted 21 cm emission from the IGM at high
redshifts. Unless the radio source population cuts off at flux levels
above the planned sensitivity of SKA, its clustering noise component
will dominate the angular fluctuations in the 21 cm signal. The
integrated foreground signal is smooth in frequency space and it
should nonetheless be possible to identify the sharp spectral feature
arising from the non-uniformities in the neutral hydrogen density
during the epoch when the first UV sources reionize the intergalactic
medium.

\end{abstract}


\section{Introduction}

Studies of Ly$\alpha$ forest in the spectra of high redshift QSOs have
now clearly established that the Universe is highly ionized by $z \sim
6$ and that the intergalactic medium (IGM) developed extensive
nonlinear structures at these high redshifts. Although it is now
possible to simulate the formation of the first ionizing objects in
the universe (Gnedin \& Ostriker~1997, Abel et al.~1998, Abel, Bryan
\& Norman~2000, Gnedin~2000, Gnedin \& Abel~2001), the epoch at which
these objects were formed, their nature and the timescale over which
the transition from the neutral to the reionized universe occurs is
currently unknown.

Building on the pioneering work of Field (1958, 1959) Scott and Rees
(1990), Madau, Meiksin and Rees (1997; MMR97 hereafter) proposed that
21 cm tomography can probe the IGM prior to the epoch of full
reionization ($z > 6$). The idea is that the radiation from the first
discrete sources should be detected indirectly through its interaction
with the surrounding neutral IGM and in the resulting emission and/or
absorption against the cosmic microwave background (CMB) at the
frequency corresponding to the redshifted 21 cm line. The resulting
patchwork (both in angle and in frequency) in the 21 cm radiation
resulting from non-uniformities in the distribution of gas density and
distribution of Ly$\alpha$ sources (i.e., the ``Cosmic web'' at these
early times) should be measurable with the next generation of radio
telescopes (e.g., Tozzi et al. 2000; hereafter T00) such as the
proposed Square Kilometer Array
(SKA)\footnote{http://www.nfra.nl/skai/} or the Low Frequency Array
(LOFAR)\footnote{http://www.astron.nl/lofar/}.

In this paper, we assess the extent to which the detection of
the redshifted 21 cm emission fluctuations is impeded by confusion
noise from extra-galactic foreground of radio sources.
 
In the following section we briefly summarize the physical origin and
expected magnitude of the 21cm signature. Section~\ref{radio} describes
the extra-galactic radio foreground emission and calculate its expected
confusion noise component. Finally, in \S 4 and \S5 we present and
discuss our results.

\section{The expected Signal}

Detailed investigations of the absorption/emission signal expected in
the 21 cm radiation have been carried out by MMR97 and T00. It is
anticipated that a strong soft UV background radiation field is
established by the first stars and quasars prior to reionization
(e.g., Haiman, Abel, and Rees 2000). Lyman alpha photons from this
background will couple the spin temperature of the neutral IGM to its
kinetic gas temperature predominantly via the Wouthuysen--Field effect
(Wouthuysen 1952; Field 1958). Since the IGM thermally decouples from
the CMB at redshift $\sim 130$ and cools adiabatically thereafter
($T_{IGM} \propto (1+z)^2$) one finds $T_{IGM}/T_{CMB} \ll 1$ before
the first luminous objects reheat the universe. Here the signal is in
absorption against the CMB, and it is expected at the wavelength of $21cm \times
(1+z_{Ly\alpha})$, where $z_{Ly\alpha}$ denotes the redshift at which
the spin temperature is first coupled to the kinetic gas temperature.
Once the IGM is heated to a temperature above that of the  CMB, the
signal will be in emission and roughly independent of the spin
temperature. The differential antenna temperature (the brightness
temperature $T_B = T_{CMB} \, e^{-\tau} + T_S\, [1-e^{-\tau}] \,$)
observed at Earth between such a patch of IGM at a spin temperature
$T_S$ and the CMB is approximately given by (T00),
\begin{eqnarray}
\left<\delta T^2 \right>^{1/2}&\approx& 10 {\rm mK}\, h^{-1} 
\left( \frac{\Omega_B h^2}{0.02} \right)
\left( \frac{1+z}{9}\right)^{1/2} \nonumber \\
& &\left( 1-\frac{T_{\rm CMB}}{T_S}\right),\label{delT}
\end{eqnarray}
where $h$ denotes the current Hubble constant in units of
$100\kmpspMpc$, and $\Omega_B$ gives the baryon density in units of
the critical density.  $|\delta T|$ is larger by a factor
$T_{CMB}/T_S$ ($\approxlt 18$ for reionization redshifts $>6$) for the
absorption signal than for emission.  Unfortunately, a relatively
large radiation flux in Ly$\alpha$ is required to ensure the coupling
of the spin temperature to the kinetic gas temperature. Therefore, the
recoil from Ly$\alpha$ scatterings most likely heats the IGM to or
above the CMB temperature before a sufficient Ly$\alpha$ background
flux is established (MMR97). Consequently, the largest signals
predicted by T00 and MMR97 are $\approxlt 10$ mK (at e.g., 150 MHz if
$z \sim 9$), as expected from equation~(\ref{delT}) and the ``Cosmic
Web'' at these times is most likely to be probed in emission at 21 cm.
The appropriate resolution for measuring fluctuations in the IGM is of
the order of 1 arc-minute with the frequency window of 1 MHz around
150 MHz (MMR97; T00).  However, if the absorption feature were
observed, it would give valuable insights to the epoch at which the
first stars formed (MMR97; T00).  At the epoch of reionization
``breakthrough'' the 21cm signal from diffuse gas decreases
steeply. One should be able to observe this signal integrated over
most of the sky even with moderate instruments (Shaver et al.~1999).

In the following sections we examine the confusion noise introduced at
$150$ MHz by extra-galactic radio sources and discuss how this might
impose serious limitations for the 21-cm tomography.

\section{Extra-galactic Radio Foregrounds}\label{radio}

SKA is planned to operate in the frequency range 0.01-20 GHz with an
angular resolution of about 1-10 arc-seconds and a sensitivity down to
a few tens of nJy.  Limitations on the sensitivity achievable for the
measurement of redshifted 21-cm signal will be set by the
contamination from the galactic and extra-galactic foregrounds.  The
dominant galactic contribution is the synchrotron background which
comprises a fraction of about 70\% at 150 MHz. On the angular scales
that are most relevant for the 21cm tomography, we nonetheless expect
the dominant angular fluctuations to be caused by clustering and
discreteness of extra-galactic sources (discussed below).

The galactic foreground could only make a comparable contribution on
sub--arc--minute scales if there were, throughout the interstellar
medium and galactic halo, fluctuations of order unity in its volume
emissivity. Such fluctuations would exist, but only in small regions
(e.g. active supernova remnants). On larger angular scales, where the
galactic foreground fluctuations would be relatively more important, it
has been sufficiently well-studied (although at much higher radio
frequencies than considered here) because of its importance for the
observations of the cosmic microwave background radiation (see
references in Tegmark \& Efstathiou~1996).  A detailed discussion of
the Galactic spectral contamination to the redshifted 21-cm is given
in Shaver et al. 1999.

Detailed multi--wavelength observations of the galactic radio emission
could be modeled sufficiently accurately for our purpose (at least in
some regions of the sky where observations would then take place).
Hence, we limit our discussion to the confusion noise introduced by
extra-galactic foreground sources such as radio galaxies, AGN and
normal galaxies which are likely to dominates the radio counts at the
low flux density levels.

\subsection{Counts of radio sources at low radio frequencies}

The wavelength region of interest is from 50 MHz ($z\sim 30$) to 200
MHz ($z\sim 6$).  To evaluate the impact of extra-galactic foreground
radio sources in this wavelength range it is necessary to model the
number density of sources as a function of flux (the differential
counts $N(S)$ per steradian). At present, the appearance of the radio
sky at flux density levels below $ 1 \mu$ Jy is not well known. However deep
VLA surveys have allowed to extend direct determinations of radio
source counts down to a few $\mu $Jy (at $\nu \approxgt$ 1.4 GHz),
implying a coverage of about 7 orders of magnitude in flux. At lower
frequencies the limiting flux densities are higher because of the confusion
noise introduced by extended sources.

The source counts from the 6C survey (Hales, Baldwin \& Warner 1988)
which was carried out at 151 MHz are a useful guide. The radio source
counts at all frequencies are typically well described by a Euclidean
power law region at the highest flux densities followed by a flatter
portion at lower flux densities (e.g. Formalont et al. 1991).  We therefore
extrapolate the 151 MHz differential source counts by a double
power-law fitted to the observed counts. This gives:

\begin{equation}
 N(S)  = \left\{
  \begin{array}{ll}
     k_1 S^{-\gamma_1}, & \hbox{$ S < S_0$} \\
   k_2 S^{-\gamma_2},   & \hbox{ $S > S_0$}   \\
  \end{array}\right.\;
\end{equation}
where $\gamma_1 = 1.75$, $\gamma_2 = 2.51$, $k_1=k_2 S_0^{\gamma_2-\gamma_1}$
with $k_2= 4.0$ per sr per mJy and $S_0 = 880 $ mJy.

This fit also includes the counts from the 3CR survey and the 3 CRR
catalogue at 178 MHz (Laing, Riley \& Longair 1983) transposed to 151
MHz assuming a mean spectral index $\alpha = 0.75$ ($ S\propto
\nu^{-\alpha}$), typical for the emission from extended lobes at these
frequencies (e.g.; Laing, Riley \& Longair 1983).  The limiting flux
density for these survey was $\sim 100$ mJy and our extrapolation is
somewhat uncertain. However, we will show that our main
conclusions are insensitive to the particular choice of the source
counts.

\subsection{Sky fluctuations from unresolved sources}

The contribution to CMB fluctuations from randomly distributed sources
of various nature has been extensively discussed in the literature
(e.g. Scheuer 1974 Cavaliere \& Setti 1976; Franceschini, Vercellone
\& Fabian 1998; Tegmark \& Efstathiou 1996; Scott \& White 1998;
Toffolatti et al. 1999; Perna \& Di Matteo 2000).  We can apply these
techniques directly to the problem at hand.  We estimate the confusion
noise due to unidentified sources below a flux density limit $S_{cut}$ which
will contribute to fluctuations in the energy band of redshifted 21-cm
emission. In the absence of clustering (Poisson distribution), the
angular power spectrum, $C_l$, is a white noise ($C_{l} \sim$
constant). The contribution to the background below the flux cut
$S_{cut}$ due to sources with a Poisson distribution is given by: \beq
C_l^{Poisson} = \langle S^2 \rangle = \int_0^{S_{cut}} S^2
\frac{dN}{dS} dS\;.  \eeq

However, the analysis of large samples of nearby radio-galaxies has
shown that sources are typically strongly clustered (e.g. Peacock \&
Nicholson 1991).  Clustering decreases the effective number of objects
in randomly distributed pixels and consequently enhances the pixel-to
pixel fluctuations (e.g.; Peebles 1980).  In the case of a powerlaw
angular correlation function ($w(\theta) = (\theta/\theta_0)^{-\beta}$
) the power spectrum of intensity fluctuations due to clustered
sources can be simply estimated as (e.g. Scott \& White 1999)
\beq
C_l^{cluster} = w_l I^2
\eeq
where $ w_l \propto l^{\beta -2}$ is the Legendre
transform of $w({\theta})$ and $I = \int_0^{S_{cut}} S (dN/dS) dS$ is
the background contributed by sources below $S_{cut}$.
If sources are clustered like galaxies today or as Lyman-break
galaxies (Giavalisco et al. 1998) at $z \sim 3$ we expect $\beta
\approx 0.8-0.9$ (most of the sources at the faint flux density
levels of interest here would in-fact be galaxies). 
The angular correlation $\theta_0$, is roughly independent of redshift
because of the bias of high redshift halos and typically of the order
of a few arc-minutes (see e.g., Oh 1999). Here we take $\theta_0 = 4$
arc-minutes.

We can compute the angular {\em rms} temperature fluctuation from 
\beq
\langle T_{rms}^2 \rangle^{1/2} = 
\left[\frac{l(l+1)C_l}{4\pi}\right]^{1/2} \left(\frac{\partial
B_\nu}{\partial T}\right)^{-1} 
\eeq
where
\beq
\frac{\partial B_\nu}{\partial T} = \frac{2k}{c^2}\left(\frac{k T}{h}\right)^2
\frac{x^4 e^x}{(e^x-1)^2} 
\label{eq:conv}
\eeq 
is the conversion factor from temperature to flux density (per steradian),
$B_\nu (T)$ is the Planck function, $x\equiv h\nu/kT$ with $T=2.725$ K
(Mather et al.~1999) the CMB temperature.  Note that the temperature
fluctuations increase with increasing angular resolution.

\section{Confusion Noise}

The $\it rms$ brightness temperature fluctuation for various values of
$S_{cut}$, as a function of angular scale $\theta$, is shown in
Figure~1. The thicker lines represent the clustering term and the
thinner lines the Poisson term.

The foreground signal from the point sources is typically larger than
the $10$mK signal (eq.~\ref{delT}) expected from the 21 cm radiation
from the IGM at high redshifts.  As expected, the signal decreases for
decreasing values of $S_{cut}$, i.e. if more sources can be identified
and removed.  In particular, as the value of the cutoff flux $S_{cut}$
is lowered, the Poisson term declines a lot more sharply than the
clustering term, until the latter totally dominates.  Given our
$dN/dS$, both components are dominated by objects just below the
detection threshold $S_{cut}$, but this dependence is stronger for the
Poisson term. Thus subtraction of bright sources decreases the Poisson
term much more effectively than the clustering term.

The SKA rms sensitivity is estimated to be (Taylor \& Brown 1999)\footnote{
See also http://www.nfra.nl/skai/science/}
\beq
S_{\rm instr} = \frac{2KT_{\rm sys}}{A_{\rm eff} \sqrt{2t\Delta \nu}}
\sim 0.3 \mu {\rm Jy} \left(\frac{1 {\rm MHz}}{\Delta \nu}
\right)^{1/2} \left(\frac{ 100 {\rm hr}}{t} \right)^{1/2}
\eeq 
at 150 MHz. Therefore, in flux density units,  the instrumental noise $S_{\rm
instr}$ is independent of angular resolution.  Given the appropriate
angular resolution required to beat down source confusion one can
therefore identify and remove foreground sources down to flux densities of
e.g. $S_{\rm cut, lim} = 7 S_{\rm instr}$. A limit on the angular resolution required
to identify all sources down to $S_{\rm cut, lim}$ can be determined
by comparing $S_{\rm cut, lim}$ to the confusion noise due to sources below 
this cutoff:
\beq
\frac{S}{N} \sim \frac{S_{\rm cut}}{(C_l^{Cluster} \|_{S_{\rm cut}} \theta^2)^{1/2}}
\sim 10 \left(\frac{S_{\rm cut}}{2.1 \mu {\rm Jy}}\right)^{0.75} \left(\frac{\theta}{0.3''}\right)^{-1.55}
\eeq
Given that SKA is planned to achieve an angular resolution of the order
of a fraction of arc-second, in principle it may be possible to remove all
sources down to $S \sim S_{\rm cut, lim}$ without being limited by the
confusion noise of the sources below that flux density.   
Figure 1 shows that even if
sources are subtracted down $S_{cut} = 1 \mu$ Jy $\sim S_{\rm cut,
lim}$, the foreground component (the Poisson one for $\theta < 10$
arc-minutes and clustering one for any $\theta$) still dominates the
primary 21 signal.

Note also that $S_{\rm cut, lim}$ will not be achieved if sources are
even slightly extended, i.e. if their angular extent $\theta \approxgt
0.3$ arc-seconds which is a small number even at large distances.  At
these faint flux density levels intrinsic source confusion due to their finite
sizes would become important (see e.g., Kellermann \& Richards 1999)

These results clearly depend on the extrapolation of the radio source
population (Eq.~2) down to flux density levels of the order of SKA
sensitivity. Clearly, if the radio source population were to cut-off
at flux densities above the SKA detection threshold, then all sources could be
identified and removed leaving no foreground contamination to the
21-cm signal.

However, this is unlikely given the typical shallow spectral
slopes of radio sources and in particular the results from the deep
VLA surveys, which found that radio counts at higher frequencies
extend all the way down to the current instrumental sensitivities
(e.g. at 5 GHz to flux densities $\sim \ 1 \mu$Jy) without any evidence of a turn over.
Indeed, VLA observations have revealed a further upturn of the
differential counts of compact sources below a few $\mu$Jy (Mitchell
\& Condon 1985; Windhorst et al. 1985) implying $N(\ge S) \propto S^{-1.2}$.

\section{Using frequency information} 

As we have seen in the previous section, within a realistic beam size
one is integrating over a very large number of unresolved
extra-galactic sources. Their individual spectra are rather shallow and
typical observed flux densities of galaxies, supernovae remnants and
the sky background etc. scale with $\nu^{-\beta}$ with $0.8 <\beta <
0.2$ (Zombeck 1982). Their integrated flux will hence also have a
rather shallow slope. 
\begin{figurehere}
\centerline{\epsfysize=3.7in\epsffile{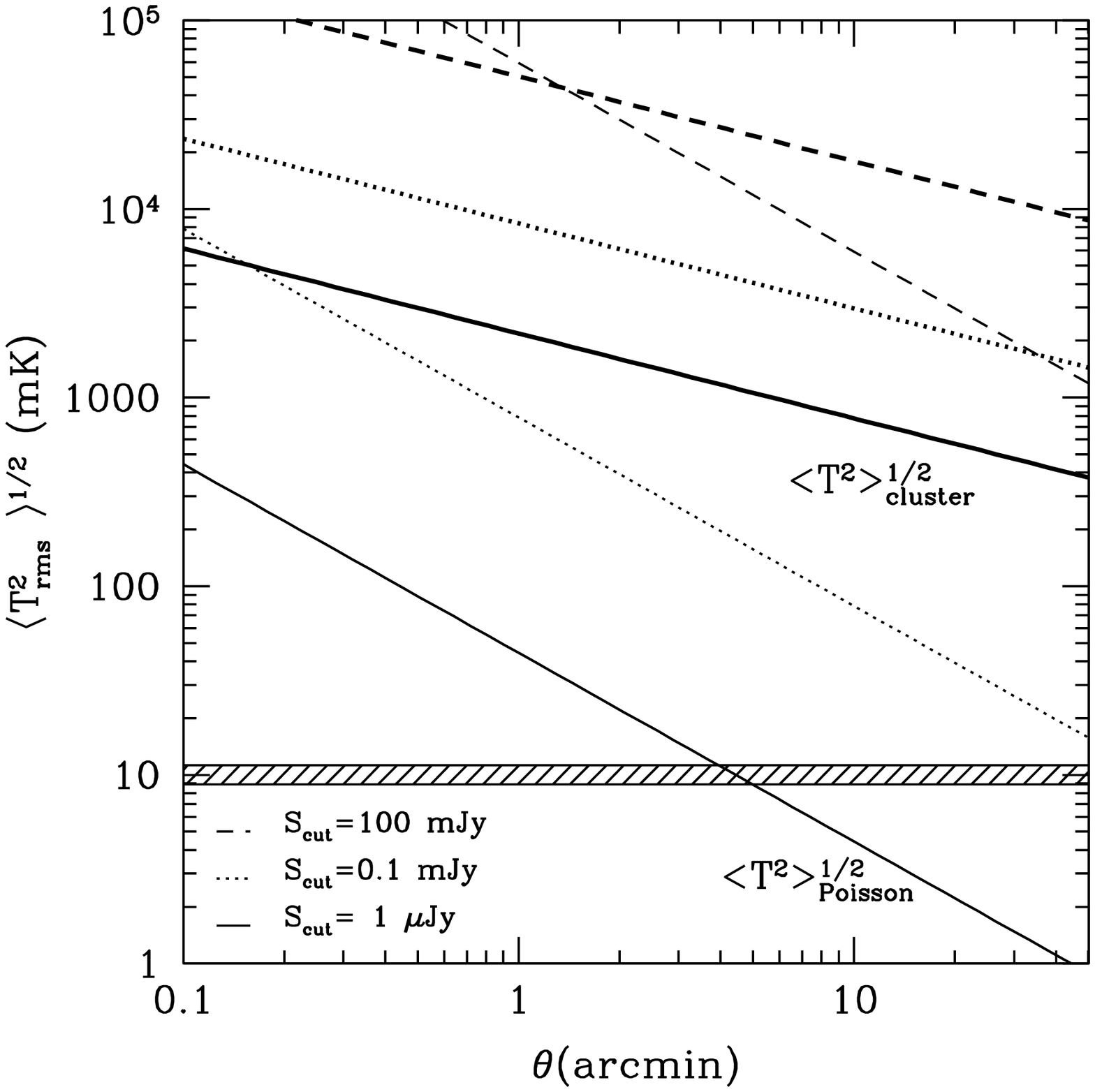}}
\caption{Angular dependence of the temperature fluctuations 
due to foreground sources at 150 MHz.  Here the signal is shown after
removal of all sources with flux densities $S>S_{\rm cut}$ as indicated in the
figure. For each value of $S_{\rm cut}$, the thin lines show the
Poisson noise, while the thick lines show the corresponding noise if
the sources are clustered.  For $S_{\rm cut}
\approxlt 0.1 $ mJy, the clustering term dominates the Poisson term at the
scales of interest for the 21cm observations.  
Note that the largest expected signal of 21cm emission at
high redshifts is below 10mK indicated by hatched region.}
\label{fig:2}\vspace{.2cm}
\end{figurehere}
The 21 cm contribution, however, has steep
spectral structures. 
The spectral index along a line of sight changes
by an amount of order unity over a small frequency range $\triangle
\nu$, where $\triangle \nu/\nu$ is the scale of neutral hydrogen
non--uniformities relative to the Hubble radius. This change may occur
multiple times along some lines of sight because of ionized regions
prior to ionization ``breakthrough''.  Subtracting out the measured
powerlaw of the foregrounds one should see the left-over frequency
dependence of the 21 cm contribution. Since the 21 cm signal is so
sharp in frequency space, small errors in the measured integrated
foreground spectrum will not hinder this measurement. For example,
observing with 2 MHz resolution with an error of $\triangle \beta\sim
.05$ of the measured foreground spectrum should allow to distinguish a
signal that is $(\nu/\nu_0)^{\triangle \beta} - 1 =
(152/150)^{\triangle \beta} - 1 \sim 6\tento{-4}$ times smaller than
the foreground. Note that this argument assumes that the beam shape
does not change for observations at a different frequency.  Obviously,
if the beam changes just slightly in size the clustering of the
unresolved foregrounds will introduce noise in the spectra on the
order of the change of area of the beam.  However, it is not clear
whether the large arrays needed to detect the signal will allow such a
good control over the synthesized beam and a more detailed study of
these technical limitations seems warranted.  If these issues could be
overcome, this signal would give very strong constraints on the nature
of the sources reionizing the intergalactic medium. Not only could one
pin down the epoch of reionization but one could map the typical size
of the regions of influence of individual sources.

\section{Conclusions}

Radio observations at meter wavelengths with instruments such as the
SKA should map the large scale structures at high redshift when the
IGM had no yet been reionized ($6\approxlt z \approxlt 30$). They
should therefore provide a useful tool for probing the epoch, nature
and sources of the reionization in the universe and their implications
for cosmology.

Assuming that the radio Galactic foreground can be sufficiently
modeled out (see Shaver et al 1999, for possible strategies to do so),
we have shown that the confusion noise provided by extra-galactic
radio sources provides a serious contamination to the brightness
fluctuations expected in the redshifted 21 cm emission from the IGM at
high redshifts. In particular, even if the radio source population is
fully identified and removed down to the flux densities corresponding to the
planned sensitivity of SKA, its clustering noise component will
totally dominate the 21 cm signal at all scales and its Poisson
component at scales $\theta < 10$ arc-seconds. Therefore, the
detection of brightness fluctuations in the redshifted 21-cm seems
unfeasible.

Note however that even in the case of significant foreground
brightness fluctuations the rise and decay of the 21 cm emission is
expected to show very sharp features in frequency space which are
potentially measurable even in the presence of the strong
extra-galactic (or galactic) continuum.  The integrated spectrum of
extra-galactic foreground sources will have a featureless powerlaw
energy spectrum which will be measured directly and can modeled out in
frequency space. To detect the spectral deviations from the redshifted
21 cm signal one would require observations in a sufficiently narrow
bandwidth $\approxlt 5$MHz.  In particular, given  such a narrow 
bandwidth, searching for the 21-cm signal should be possible
even if there are spectral variations in the foregrounds.  Any
variation in frequency space in the foregrounds would be correlated
(as the galactic or extragalactic foreground have power law spectra)
whereas the 21 cm feature would appear as an uncorrelated signal and
should therefore be detectable.
Technical challenges as e.g.  techniques that minimize the change in
the effective beam size as one changes the frequency band will have to
be addressed.

\acknowledgments 

T.\,D.\,M.\ acknowledges support for this work provided by NASA
through Chandra Postdoctoral Fellowship grant number PF8-10005 awarded
by the Chandra Science Center, which is operated by the Smithsonian
Astrophysical Observatory for NASA under contract NAS8-39073.
R.P. aknowlodges support from Harvard Society of Fellows.
T.A. acknowledges support from NSF grants ACI96-19019 and AST-9803137
and is grateful to many insightful discussion with Torsten En{\ss}lin
and Andreas Quirrenbach on radio observations.


\begin{references}
\reference{} Abel, T., Bryan, G.\ L.\ \& Norman, M.\ L.\ 2000, \apj, 540, 39 

\reference{} Abel, T., Norman, M.\ L.\ \& Madau, P.\ 1999, \apj, 523, 66 

\reference{} Cavaliere, A. \& Setti, G. 1976, A\&A, 46, 81

\reference{} Condon, J. J. 1974, ApJ, 188, 279

\reference{} Field, G. B. 1958, Proc. IRE, 46, 240 

\reference{} Field, G. B. 1959, \apj, 129, 536

\reference{} Franceschini, A., Vercellone, S., \& Fabian, A. C. 1998, MNRAS, 29, 817

\reference{} Formalont, E.B., Windhorst R.A., Kristian J.A., Kellerman K.I., 1991, AJ, 102, 1258

\reference{} Gnedin, N.\ Y.\ 2000, \apj, 535, 530 

\reference{} Gnedin, N.\ Y.\ \& Abel, T. 2001, NewA, in press

\reference{} Gnedin, N.\ Y.\ \& Ostriker, J.\ P.\ 1997, \apj, 486, 581 

\reference{} Haiman, Z., Abel, T. \& Rees, M. J. 2000, ApJ, 534, 11

\reference{} Hales, S.\ E.\ G., Baldwin, J.\ E.\ \& Warner, P.\ J.\ 1988, MNRAS, 234, 919

\reference{} Kellermann, K.\ I., Richards, E.\ A., 1999, to appear in 'Scientific Imperatives at Centimeter Wavelengths,' eds. M. P. van Haarlem and J. M. van der Hulst 1999, Dwingeloo, NFRA (astro-ph/9909083) 

\reference{} Laing, R.\ A., Riley, J.\ M.\ \& Longair, M.\ S.\ 1983, MNRAS, 204, 151 

\reference{} Madau, P., Meiksin, A.\ \& Rees, M.\ J.\ 1997, ApJ, 475,
429, (MMR97)

\reference{} Mather, J.\ C., Fixsen, D.\ J., Shafer, R.\ A., Mosier, C., \& Wilkinson, D.\ T.\ 1999, \apj, 512, 511

\reference{} Mitchell, K.\ J.\ \& Condon, J.\ J.\ 1985, AJ, 90, 1957 

\reference{} Perna, R., \& Di Matteo, T. 2000, ApJ, 542, 68

\reference{} Peebles P.J.E., 1993, Principles of Physical Cosmology (Princeton: Princeton University Press)

\reference{} Peacock J.A., Nicholson S.F., 1991, MNRAS, 253, 307

\reference{} Razoumov, A.\ O.\ \& Scott, D.\ 1999, \mnras, 309, 287 

\reference{} Shaver, P. A., Windhorst, R. A., Madau, P., \& de Bruyn, A. G. 1999, A\&A, 345, 380

\reference{} Scheuer, P. A. G. 1974, MNRAS, 166, 329

\reference{} Scott, D.\ \& Rees, M.\ J.\ 1990, MNRAS, 247, 510 

\reference{} Scott, D., White M., 1999, A\&A, 346, 1

\reference{} Tegmark, M. \& Efstathiou, G. 1996, MNRAS, 281, 1297

\reference{} Toffolatti, L., Gomez, F. A., De Zotti, G., Mazzei, P.,
Franceschini, A., Danese, L. \& Burigana, C. 1999a, MNRAS, 297, 117

\reference{} Tozzi, P., Madau, P., Meiksin, A.\ \& Rees, M.\ J.\ 2000,
\apj, 528, 597, (T00)

\reference{} Windhorst, R.\ A., Miley, G.\ K., Owen, F.\ N., Kron, R.\ G.\ \& Koo, 
D.\ C.\ 1985, ApJ, 289, 494 

\reference{} Wouthuysen, S. A. 1952, AJ, 57, 31 

\reference{} Zombeck, M. V. 1982, Handbook of Space Astronomy and
Astrophysics, Cambridge University Press

\end{references}
\end{document}